# The equation of state of neutron stars and the role of nuclear experiments

Francesca Gulminelli[1] and Anthea F. Fantina[2]

(1) Normandie Univ., ENSICAEN, UNICAEN, CNRS/IN2P3, LPC Caen, F-14000 Caen, France.
(2) Grand Accélérateur National d'Ions Lourds (GANIL), CEA/DRF-CNRS/IN2P3, Bvd. Henri Becquerel, 14076 Caen, France.

## Introduction

Neutron stars (NSs) are unique laboratories to probe matter in extreme conditions, not accessible in terrestrial laboratories. Indeed, with a mass of about 1-2 solar masses and a radius of about 10 km, NSs have densities spanning several orders of magnitude, exceeding in their core the density found inside atomic nuclei. While the properties of the outer layers consisting of a (solid) crust of nuclei embedded in an electron gas, together with a neutron gas in the denser region (possibly on top of a nuclear "pasta" mantle), and those of the outer core made of homogeneous asymmetric nuclear matter, can, to some extent, be constrained by nuclear physics experiments and theory, the nature of the NS inner core formed of strongly interacting matter at very high density, still remains largely unknown [1].

The remarkable advances of multi-messenger astronomy on different dense-matter astrophysical sources has very recently led to the first quantitative measurements of various properties of NSs such as the mass-radius relation [2-4] and the tidal polarizability [5].
These new observations, as well as the plethora of upcoming results from the LIGO/Virgo collaboration and those expected from the next generation of detectors, raise hopes in the next future to answer fundamental open questions such as the structure and degrees of freedom of baryonic matter in extreme density conditions, the occurrence of phase transitions, and the presence of deconfined matter in the core of NSs. In turn, these new powerful probes of ultra-dense matter have boosted the experimental and theoretical research on nuclear matter, and specifically the study of the nuclear-matter equation of state (EoS), that is, the functional relation $P(\rho_B)$ between pressure and mass-energy density of baryonic matter.

This direct connection between astrophysical measurements and the microphysics of dense matter is due to the fact that, under the unique hypothesis of validity of general relativity, there is a one-to-one correspondence between any static observable of mature and slowly rotating NSs and the underlying EoS. In particular, the Tolman-Oppenheimer-Volkoff (TOV) equations of hydrostatic equilibrium allow to associate the EoS to a unique relation between the mass $M$ and the radius $R$ of the NS [1]. This means that simultaneous measurements of mass and radius of different NSs would give a direct measure of the pressure of baryonic matter at given densities (more massive stars being associated to higher densities), thus a measurement of the EoS.

Unfortunately, the most precise measurements of NS masses concern binary systems, while radii have been recently inferred from the pulse profile modelling of X-ray data from (isolated) millisecond pulsars, making the observational mass-radius correlation relatively loose.

An especially interesting complementary observable is given by the tidal polarizability $\Lambda$, which measures the deformability of a compact object under the gravitational influence of a second body. The bijective correspondence between the EoS and the variation of the tidal polarizability with the star mass is known since the sixties, and a first measure of this parameter was provided by the LIGO/Virgo collaboration in the celebrated gravitational-wave event GW170817 [5].

Modelling the neutron-star equation of state

The extraction of the dense-matter composition and degrees of freedom is complicated by the fact that there is no ab-initio calculation of dense matter in the nucleonic nor in the hadronic or partonic sectors, and effective models must be used instead (see [6] for a review). In order to extract the nuclear EoS from the gravitational waveforms, agnostic inference is therefore used such as to explore all the possible shapes of the $P(\rho_B)$ functional [7]. The standard method consists in parametrizing the EoS with piecewise polytropes, subject to general physics constraints such as causality, the respect of the unitary limit, and the asymptotic properties of lattice QCD. Many alternative powerful techniques have also been proposed, such as spectral function expansions, parametrized sound speed models, as well as non-parametric inference based on gaussian processes ([7] and references therein). The domain is evolving fast, and this collective effort is leading to a model-independent prediction of the different NS properties, based on constraints given by the various astrophysical observations.

However, this considerable progress is not sufficient to pin down the microscopic structure and properties of the ultra-dense baryonic matter composing the NS.
Indeed, whereas a given macroscopic $M(R)$ or $\Lambda(M)$ curve is univocally associated to a microscopic $P(\rho_B)$ functional, identical EoSs can be obtained under different hypotheses on the underlying microphysics. Indeed, whatever the effective model employed, the energy density of baryonic matter depends on the different conserved charges of the strong interactions, namely the baryonic, charge, and strange density. On the other hand, the gravitational pressure only imposes the total energy density. The relative proportion of the different densities is determined by weak interactions, which are in equilibrium inside the NS, but the condition of chemical equilibrium explicitly depends on the (model dependent) energy functional.

Therefore, a way to extract information on the structure and properties of dense matter is to use EoS parametrizations that cover the parameter space of effective nuclear models, under the different hypotheses on the effective degrees of freedom at high density. Presently, the tighter constraints are obtained in the case of purely nucleonic models, that are routinely optimised and calibrated over a plethora of experimental nuclear data.
Parametrizing the EoS with a functional that can explore the complete parameter space of nucleonic EoSs is known as nucleonic meta-modelling [8]. It consists in introducing a flexible energy functional able to reproduce existing effective nucleonic models as well as interpolate between them. The parameter space being regulated by our present theoretical and experimental knowledge of nuclear physics, this technique allows to predict the astrophysical observables with controlled uncertainties. Moreover, it can be used as a null hypothesis to infer from the astrophysical data the presence of exotic non-nucleonic degrees of freedom. Nucleonic meta-modelling consists in a density or Fermi momentum expansion of the energy per particle of infinite nuclear matter composed of neutrons and protons with respective densities $n_n$ and $n_p$.

In particular, a Taylor expansion around the saturation density $n_0$ allows the separation of the low-order derivatives, which are better determined by nuclear theory and experiments, from the high-order ones, which contain the highest uncertainties [9]. Schematically, we can write, at the order N:

$$e^N(n_n, n_p) = t^{FG} + \sum_{k=0}^{N} \frac{1}{k!}(v_k^{is} + v_k^{iv}\delta^2)\left(\frac{n-n_0}{3n_0}\right)^k, \qquad (1)$$

where we have singled out the dominant degenerate Fermi gas contribution $t^{FG}(n_n, n_p)$, and defined the total baryonic density $n = n_n + n_p$ and the neutron-proton asymmetry $\delta = \frac{n_n - n_p}{n}$. The isoscalar (isovector) coefficients $v_k^{is}$ ($v_k^{iv}$) can be one-to-one connected to the so-called empirical parameters [6,9], given by the successive derivatives of the energy functional at saturation:

$$X_k^{is(iv)} = \left.\frac{\partial^k e_{0(sym)}^N}{\partial n^k}\right|_{n=n_0, \delta=0}, \qquad (2)$$

where $e_0^N$ is the energy of symmetric $n_n = n_p$ matter, and the density-dependent symmetry energy is defined as $e_{sym}^N = \left.\partial^2 e^N / \partial \delta^k\right|_{\delta=0}$. For realistic applications, the expansion Eq. (1) is slightly complexified, introducing the extra density dependence induced by the effective mass, a low-density correction enforcing the zero-density limit [9], and a specific treatment of the NS crust [10]. This ensures a unified EoS treatment, meaning that the same nuclear model is applied in the different regions of the NS, which is crucial to avoid ad-hoc EoS matching that may introduce unphysical effects in the NS modelling [11].

The low-order empirical parameters (up to *k*=2) are relatively well constrained by ab-initio nuclear theory and low-energy nuclear experiments. On the theory side, an impressive progress has been made in the chiral perturbation theory of nuclear interactions, with a diagrammatic expansion which allows a reliable estimation of the uncertainties due to the missing terms [12]. On the experimental side, the low-order empirical parameters have been extracted by a systematic comparison of density functional approaches with a large set of high-precision nuclear observables such as masses, skins, electric dipole polarizability, isobaric analog states systematics, flows in heavy-ion collisions, collective modes (see [6,9,13] for a review, and references therein).

A (non-exhaustive) compilation of these different constraints is reported in Fig. 1, in what concerns the symmetry energy at saturation $X_0^{iv} = J$ and its slope $X_1^{iv} = L$. The tighter constraints are provided by nuclear theory, but it is interesting to observe that nuclear phenomenology gives independent uncertainty intervals for the empirical parameters which agree with the ab-initio modelling. Among the different parameter evaluations reported in Fig.1, the only one which is in principle fully model-independent is the one based on the parity-violation electron scattering measurement of the $^{208}$Pb skin by the PREX experiment [14]. This measurement seems to suggest values of *J* and *L* considerably higher than the other estimations. However, the systematic errors are so large that no clear conclusion can be drawn at this stage.

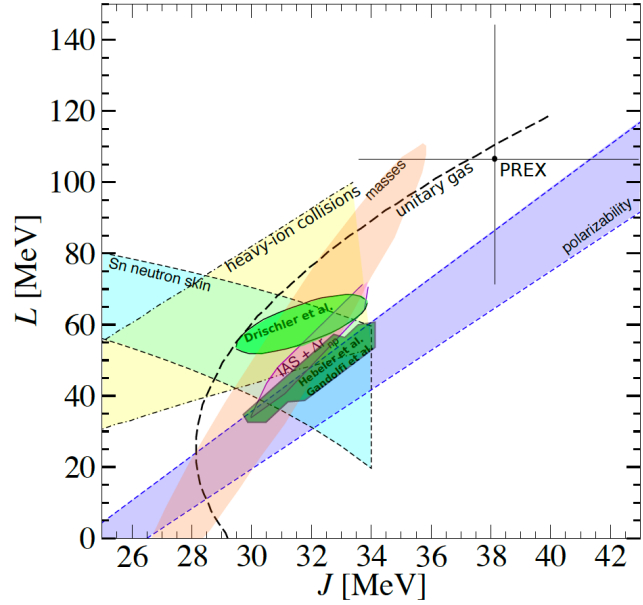

**Figure 1.** Slope of the symmetry energy *L* versus the symmetry energy coefficient *J*, as extracted from different experimental and microscopic ab-initio theoretical constraints (see [6] for original references of the constraints).

If chiral effective field theory is today the most powerful constraint on the neutron-matter EoS, the role is inverted concerning the empirical parameters $X_k^{is}$ characterizing symmetric matter. For those parameters, the precision of EoS measurement from the combined analysis of different nuclear experiments is remarkable. This is due to the fact that only nuclei between driplines can be probed in terrestrial experiments, and the properties of extremely neutron-rich nuclei are not yet fully explored. The lowest-order parameters, namely the saturation energy $X_0^{is} \equiv E_0$ and the associated saturation density $n_o$, are very precisely measured at the 2% level by nuclear mass and radii measurements, and the compressibility $X_2^{is} \equiv K_0$ is presently extracted at the 10 % level from refined analyses of giant monopole resonance data.

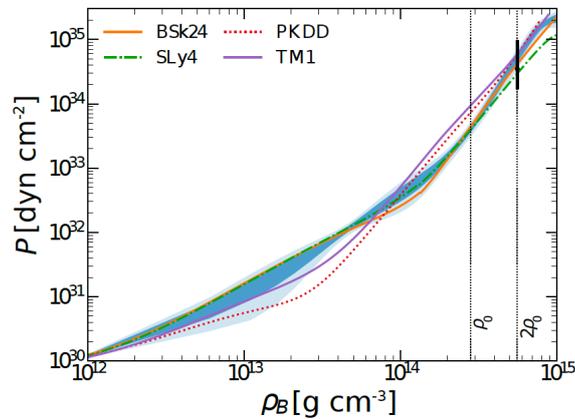

**Figure 2.** Marginalized posterior for the pressure P versus baryon mass-energy density $\rho_B$. The blue dark (light) shaded region shows the 1σ (2σ) confidence intervals. Some popular nucleonic models are represented as lines. The vertical black bar corresponds to the constraint inferred from GW170817 in [5]. Figure adapted from [10].

All these different constraints can be incorporated in a nucleonic meta-modelling representation of the nuclear EoS such as the one of Eq. (1) with the standard tools of Bayesian inference [9,10]. To ensure that models are physically viable up to the extreme densities in the core of NSs, additional requirements are considered, namely the causality of the EoS and its capability of sustaining massive NSs (above 1.97 $M_{sun}$) like the precisely measured pulsar J0348+0432. The resulting most general EoS compatible with these constraints is represented in Fig. 2. This estimation agrees with the constraint extracted from an agnostic analysis of the gravitational-wave event GW170817 [5], also reported in Fig.2. This shows a nice convergence between our micro- and macro-physics understanding of compact objects. We can also see that the knowledge on nuclear matter that can be extracted from laboratory experiments and ab-initio nuclear theory is very compelling, and it could be worthwhile to include it in future astrophysical data analyses to improve their predictive power. Some first studies, where the nuclear physics constraints are directly included in the data analyses, start to be proposed [15]. It is also interesting to observe that several popular models currently used to interpret nuclear data fall outside the range of the posterior estimation of the pressure in Fig. 2. This is because most nuclear models have been optimised to reproduce specific properties of nuclear data, notably in nuclear structure, where a very limited range of densities and isospin asymmetries are explored. Therefore, their extrapolation to beta-equilibrated very neutron-rich matter in a large density domain might not be compatible with the combined set of constraints.

Are non-nucleonic degrees of freedom excluded?

The unique hypothesis of the nucleonic meta-modelling is that the relevant degrees of freedom inside the core of a NS are given by protons and neutrons (supplemented by electrons and muons). In this respect, the EoS estimation of Fig.2 can be considered reliable, since it is reasonable to suppose that baryonic matter should be purely nucleonic up to around twice the saturation density. Despite considerable uncertainty exists about the possible presence of hyperons in the core of NSs, the high mass of the lightest lambda hyperon imposes that the threshold density for its possible appearance be above about $2n_0$ [16].
Concerning the possible presence of deconfined matter, agnostic modelling of a first-order phase transition including constraints from LIGO/Virgo and NICER data seems to exclude the possibility of a transition below $\sim 2n_0$ [17]. If we believe our nuclear physics constraints, the only possible scenario allowing a violation of the EoS in Fig.1 would be the absolute stability of quark matter, which, if true, would lead to the existence of pure quark stars [1,18].
This shows the huge challenge connected to an increased precision of the measurement of the NS properties at moderate densities from future astrophysical observations.

The situation is different when we analyse observables that imply an integration over the whole density profile of the NS. For the mass-radius relation (Fig. 3, upper panel), the parameter space compatible with nucleonic degrees of freedom is large, because densities as high as $\sim 6-8\, n_0$ can be reached in the core of the most massive NSs, and the associated EoS is dominated by the high-order empirical parameters $X_k^{is(iv)}$ (with k>2). Those parameters are virtually unconstrained by nuclear physics experiments, which probe densities close to saturation where their influence is too small and low-order parameters dominate. Concerning ab-initio nuclear theory, the power counting associated to the diagrammatic expansion breaks down when the short-distance structure cannot be neglected, thus those calculations, though extremely powerful at low density, cannot be safely extrapolated beyond $\sim 1.5\, n_0$ [12]. These limitations of both nuclear theory and experiments lead to an increased uncertainty in the EoS at high density. Despite these large uncertainties, the nuclear physics prediction of the *M(R)* relation is tighter than the present astrophysical measurements (see Fig. 3). Combining the whole set of experimental and theoretical filters is definitely important for a reliable prediction of astrophysical observables.

In the case of the *M(R)* relation, the hypothesis of nucleonic degrees of freedom implicit in the meta-modelling is certainly restrictive. Indeed, even if the exact location of the deconfinement transition towards quark matter is essentially unknown, if strange quark matter is not absolutely stable such a transition is expected at high density [18]. In particular, if this transition is strongly first order, the parameter space of the EoS increases, leading to a much larger spread of the possible *M(R)* relation ([19] and references therein).

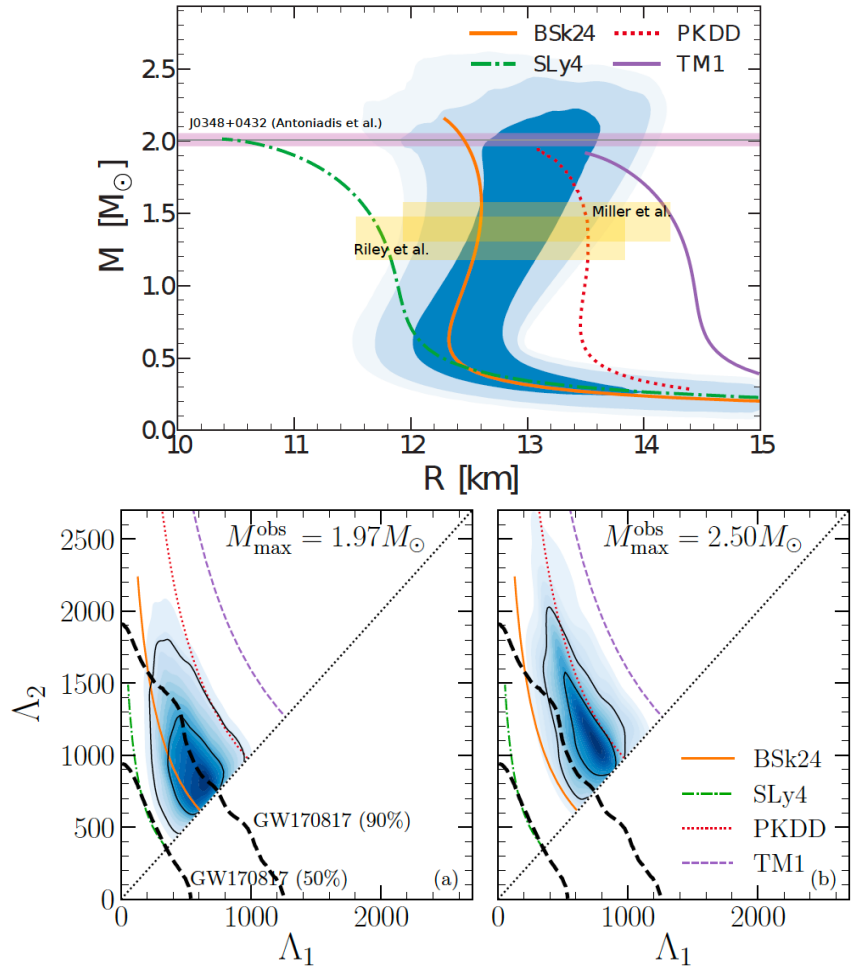

Figure 3. Upper part: marginalized posterior for the NS mass-radius relation. The blue shaded regions show the 68%, 95% and 99% confidence intervals. Lower part: Marginalized posterior for the tidal polarizability parameter of the two components of GW170817. Dark (light) shaded regions represent the 50% (90%) confidence interval with the Bayesian estimation; black dashed lines show the confidence regions as obtained in [5]. The left and right panels correspond to different constraints on the maximal NS mass. The results for the same popular EoS models as in Fig.2 are also represented. Figures adapted from [10].

Because of the phenomenological nature of the effective models used for the quark phase, settling the parameter space associated to a phase transition to quark matter is a very hard task. However, the parameter space associated to the absence of such a transition is much more limited (see Fig. 3). This means that the nucleonic meta-modelling can be used as a null hypothesis to set the existence of deconfined matter in the core of NSs, provided that the observations are sufficiently precise to challenge the nucleonic predictions.

The first steps in this direction come from the recent measurement of the tidal polarizability parameter $\Lambda$ in compact-binary mergers [5]. The compatibility of the posterior distribution of the $\Lambda$ parameter shown in Fig. 3 (lower left panel) with the LIGO/Virgo gravitational-wave measurement at the 90% level indicates that there is no compelling evidence that exotic matter is present in the NS core. However, the shape of the probability distribution predicted by low-energy nuclear physics constraints is very different from the one inferred from the GW analysis. New astronomical observations will hopefully produce more stringent constraints in the next future. For example, if a NS as massive as 2.5 $M_{sun}$ were measured, this would lead to a clear incompatibility between the $\Lambda$ measurement and the purely nucleonic hypothesis, as shown in Fig. 3 (lower right panel), thus providing a more convincing evidence of the presence of deconfined matter in compact stars.

### Future constraints from nuclear-physics experiments

In the absence of such astrophysical evidence, an alternative path to challenge the nucleonic hypothesis would consist in putting more stringent constraints on the nucleonic meta-modelling prediction from improved laboratory measurements. A reduced error bar in the neutron skin measurement such as that performed by the PREX collaboration [14] would greatly constrain the low-order empirical parameters in the isospin sector, thus allowing better extrapolations to higher density, and reduced uncertainties in the astrophysical predictions. The same is true for measurements from relativistic heavy-ion collisions, which are a unique experimental probe of densities of the order of $2n_0$. Only symmetric matter is probed in such experiments, but even a loose constraint on the higher-order parameter $X_3^{is} \equiv Q_0$ would greatly reduce the present uncertainty in the nucleonic extrapolation of the high-density EoS. Encouraging results in this sense have been recently presented by the HADES collaboration [20].

To conclude, even if the composition, properties and degrees of freedom of ultra-dense matter such as can be found in the core of NSs is still largely unknown, we are presently facing an extremely exciting and rapidly evolving scientific epoch. For the first time, terrestrial nuclear experiments can be connected to astronomical observations, and it makes no doubt that, together with the amazing recent progress of ab-initio theoretical modelling, they will further contribute in a next future to unveil the remaining mysteries in the properties of matter under extreme conditions.